\documentclass[prb,nofootinbib,twocolumn,superscriptaddress]{revtex4} 

\usepackage{graphicx}
\usepackage{dcolumn}
\usepackage{bm}
\usepackage{threeparttable}
\usepackage{times}
\usepackage{mathptmx}
\usepackage{lscape}
\usepackage{natbib}
\usepackage{amsmath}
\usepackage{amssymb}
\usepackage{braket}
\usepackage{comment}
\usepackage{color}

\def\degree{\kern-.2em\r{}\kern-.3em}

\begin{document}


\title{  Bidirectional-Stability Breaking in Thermodynamic Average for Classical Discrete Systems  }

\author{Koretaka Yuge}
\affiliation{
Department of Materials Science and Engineering,  Kyoto University, Sakyo, Kyoto 606-8501, Japan\\
}%

\author{Shouno Ohta}
\affiliation{
Department of Materials Science and Engineering,  Kyoto University, Sakyo, Kyoto 606-8501, Japan\\
}%

\begin{abstract}
{
For classical systems, expectation value of macroscopic property in equilibrium state can be typically provided through thermodynamic (so-called canonical) average, where summation is taken over possible states in phase space (or in crystalline solids, it is typically approximated on cofiguration space). 
Although we have a number of theoretical approaches enabling to quantitatively estimate equilibrium properties by applying given potential energy surface (PES) to the thermodynamic average, it is generally unclear whether PES can be stablly, inversely determined from a given set of properties.
This essentially comes from the fact that bidirectional stability characters of thermodynamic average for classical system is not sufficiently understood so far. 
Our recent study reveals that for classical discrete system, this property for a set of microscopic states satisfying special condition can be well-characterized by a newly-introduced concept of "anharmonicity in the structural degree of freedom" of $D$, where these states are expected to be stably inversed to underlying PES,  known without any thermodynamic information.
However, it is still quantitatively unclear how the bidirectional stability character is broken inside the configuration space.
Here we show that the breaking in bidirectional stability for thermodynamic average is quantitatively formulated:  
We find that the breaking is mainly dominated by the sum of divergence and Jacobian of vector field $D$ in configuration space, which can be fully known \textit{a priori} only from geometric information of underlying lattice,  without using any thermodynamic information such as energy or temperature.  
  }
\end{abstract}


\maketitle

\section{Introduction}
Statistical mechanics provides us that when potential energy surface (PES) of the system is once given, macroscopic properties (especiall, dynamical variables) in thermodynamically equilibrium state can be reasonablly determined through thermodynamic (so-called canonical) average, whose summation is in principle taken over all possible microscopic states on phase space.
Since number of possibles states considered astronomically increases with increase of the system size and/or the number of components, a variety of theoretical techniques have been developed to effectively predict macroscopic properties such as Metropolis algorism, entropic sampling and Wang-Landau sampling.\cite{mc1,mc2,mc3,wl} 
Despite many successes in the current classical statistical mechanics, it is still generally unclear how the canonical average can be \textit{stably bidirectional} between thermodynamically equilibrium structure and underlying PES.

We recently find that the thermodynamic average can expected to be \textit{partially}, bidirectionally stable for a set of microscopic state satisfying special condition on configurational geometry, which is characterized by a newly-introduced concept of "anharmonicity in the structural degree of freedom" of $D$ known \textit{a priori} without requiring any thermodynamic information.\cite{spe} 
Here we show that breaking in bidirectional stability of the thermodynamic average is further investigated, by extending our previous concept of the anharmonicity. We quantitatively derive the scaling factor of the breaking based on divergence and Jacobian of vector field $D$, which roughly indicates that sum of linear and nonlinear change in nonlinear contribution of thermodynamic average essentially enhance the bidirectional stability breaking. The details are shown below.

\section{Derivation and Discussions}
\subsection{ Preparation for derivation }
Before formulating the stability breaking, we first briefly explain the basic concept of the anharmonicity in the structral degree of freedom, $D$, which will play central role in the following discussions. 
Recently, we find that for a broad class of classical systems (e.g., representative lattices such as fcc, bcc, square and triangle, and liquids in rigid box), configurational density of states (CDOS) for non-interacting system can be well-characterized by multidimensional gaussian when system size increases.\cite{em1,em3} We have confirmed this character not only by comparing landscape of the practical CDOS with ideal gaussian, but also by comparing even-order moments of CDOS with those of gaussian.\cite{cm} 
We also found that when the CDOS is exactly given by multidimensional gaussian (hereinafter, we call such ideal system as "harmonic system"), thermodynamic average $\phi_{\textrm{th}}\left( \beta \right)$ for structure becomes bijective to underlying PES\cite{em2,spe} ($\beta=\left( k_{\textrm{B}}T \right)^{-1}$, where $k_{\textrm{B}}$ denotes Boltzmann constant and $T$ is absolute temperature).
Here, thermodynamic average is explicitly given by linear map, i.e., by matrix form $\Lambda$ acting on PES, whose elements linearly depends on $\beta$ and geometric information about underlying lattice. 
Using these facts, our recent study introduce \textit{anhamonicity in the structural degree of freedom},\cite{spe} $D$, defined as 
\begin{eqnarray}
D\left( Q \right) = d\left( Q, \left( \phi_{\textrm{th}}\left( \beta \right)\circ\Lambda^{-1}\left( \beta \right) \right)\cdot Q \right),
\end{eqnarray}
where $d\left( \quad, \quad\right)$ denotes distance function under Euclidean metric, and $\Lambda$ is given by\cite{em2}
\begin{eqnarray}
\label{eq:emrs}
\Lambda_{ik}\left(\beta\right) = -\beta \Braket{q_i \cdot q_k}, 
\end{eqnarray}
where $\Braket{\quad}$ denotes linear average over all possible microscopic structures: The most important point here is that the averages are calculated for a \textit{non-interacting} system. This directly means that we can construct the matrix $\bm{\Lambda}$ \textit{a priori} without requiring any information about the many-body interactions. Moreover, we have shown that image of the composite map, $\phi_{\textrm{th}}\left( \beta \right)\circ\Lambda^{-1}\left( \beta \right) $, is exactly independent of $\beta$. 
These facts means that anharmonicity $D$ is a measure of nonlinear character for thermodynamic average (because for linear map, $D=0$), which purely reflects the geometric character of underlying lattice without requiring any thermodynamic information. Hereinafter, for simplicity, microscopic structure is described measured from that at center of gravity in CDOS. 
Although we have qualitatively predict that microscopic structure with smaller magnitude of $D$ have more bidirectional-stability, it is still unclear how the stability is formulated by $D$. 

With the above preparation, we here consider classical discrete system under \textit{constant composition}, which typically refers to substitutional multicomponent alloys. Microscopic states on given lattice, $Q$, is described in terms of the corresponding coordination $\left\{ q_1,\cdots, q_f \right\}$ where $f$ denotes dimension of configuration space considered. We here employ coordination based on generalised Ising model,\cite{ce} providing complete orthonormal basis functions. Under these  definition, potential energy $U$ for any microscopic state is exactly given by 
\begin{eqnarray}
\label{eq:u}
U_p\left( Q \right) = \sum_{i=1}^f \Braket{U_p|q_i} q_i\left( Q \right),
\end{eqnarray}
where $\Braket{\quad|\quad}$ denotes the inner product, i.e. trace over all possible microscopic states, and subscript $p$ denotes the choice of landscape of the PES. 
The present study describe PES in terms of the above given coordinates $\left\{q_1,\cdots , q_f\right\}$, in inner product form, i.e., $U_p=\left( \Braket{U_p|q_1},\cdots, \Braket{U_p|q_f} \right)$.

\subsection{Derivation of Bidirectional-Stability Breaking}

\begin{figure}[h]
\begin{center}
\includegraphics[width=0.5\linewidth]
{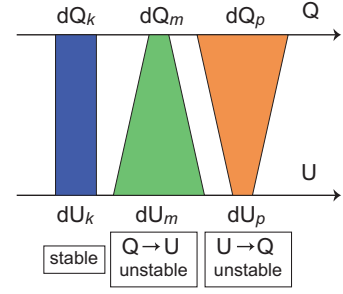}
\caption{Schematic illustration of bidiretional stability relationships in terms of ratio of hypervolume $\mathbf{dU}$ and $\mathbf{dQ}$. }
\label{fig:map}
\end{center}
\end{figure}

The bidirectional stability breaking, $B$, can be naturally defined based on the ratio of hypervolume $\mathbf{dU}^{\left(m\right)}=du_{1}^{\left(m\right)}\cdots du_{f}^{\left(m\right)}$ and $\mathbf{dQ}^{\left(m\right)}=dq_{1}^{\left(m\right)}\cdots dq_{f}^{\left(m\right)}$, where PES in region $\mathbf{dU}^{\left(m\right)}$ is mapped onto region of microscpic structure $\mathbf{dQ}^{\left(m\right)}$ through thermodynamic (canonical) average, $\phi_{\textrm{th}}$: 
\begin{eqnarray}
R^{\left(m\right)} = \frac{\mathbf{dQ}^{\left(m\right)}}{\mathbf{dU}^{\left(m\right)}}. 
\end{eqnarray}
Figure~\ref{fig:map} shows schematic illustration of bidirectional stability in terms of ratio of hypervolume $\mathbf{dU}$ and $\mathbf{dQ}$. 
The figure qualitatively indicates that bidirectional stability is satisfied for $R^{\left(m\right)} \simeq \alpha$, $Q\to U$ is unstable with $R^{\left(m\right)} < \alpha$ and $U\to Q$ is unstable with $R^{\left(m\right)} > \alpha$, where $\alpha$ is a certain constant:  
Here, the problem is how to determine $\alpha$, providing stablity condition.  
Since for $U^{\left(0\right)}=\left(0,0,\cdots, 0\right)$ (i.e., non-interacting system), corresponding canonica average should results in $Q^{\left(0\right)}=\left(0,0,\cdots,0\right)$, we start from the condition that $\mathbf{dQ}^{\left(0\right)}/\mathbf{dU}^{\left(0\right)}$ provides bidirectional stability. With this consideration, the stability breaking at microscopic structure $m$, $Q^{\left(m\right)}$ can be naturally defined as 
\begin{eqnarray}
B^{\left(m\right)} = \log \left(\frac{R^{\left(m\right)}}{R^{\left(0\right)}}\right),
\end{eqnarray} 
which means that $B=0$ is stable, $B > 0$ is $U\to Q$ unstable, and $B < 0$ is $Q\to U$ unstable.

To quantitatively formulate $B$ for individual microscopic structure, we first rewrite canonical average of structure along coordination $i$ in Taylor expansion form:
\begin{eqnarray}
\label{eq:Q}
\label{eq:taylor}
Q_i\left(\beta\right) \simeq -\sum_{j=1}^f u_j r_{ij} + \frac{1}{2} \sum_{j=1}^f \sum_{k=1}^f u_j u_k d_{ijk},
\end{eqnarray}
where
\begin{eqnarray}
u_j &=& \beta \Braket{U|q_j} \nonumber \\
r_{ij} &=& \Braket{q_i q_j} \nonumber \\
d_{ijk}&=& \Braket{q_i q_j q_k},
\end{eqnarray}
and $\beta$ is inverse temperature. We previously show that when CDOS \textit{before} applying many-body interaction to the system is exactly identical to multidimensional gaussian, 
r.h.s. of Eq.~\eqref{eq:Q} excluding the second term becomes exactly identical to the thermodynamic average, corresponding to the linear map $\Lambda$ where $\Lambda^{-1}$ always exists. 
Therefore, the second term of Eq.~\eqref{eq:Q}, corresponding to including multivariate third-order moment of CDOS, represents nonlinear contribution of theromdynamic average. 
We here focus on that the following is satisfied:
\begin{eqnarray}
R^{\left(m\right)} = \left| J^{\left(m\right)}\left( \frac{ \partial \left(Q_1,\cdots,Q_f\right)}{\partial\left(u_1,\cdots, u_f\right)} \right)\right|,
\end{eqnarray}
where $J^{\left(m\right)}$ is Jacobian at $Q^{\left(m\right)}$. 
This corresponds to that when thermodynamic average $\phi$ is a linear map w.r.t. $\left\{u_i\right\}$, we find that 
\begin{eqnarray}
\label{eq:J}
\forall m:\quad J^{\left(m\right)} = J^{\left(0\right)} = \mathrm{const.},
\end{eqnarray}
which means that bidirectional stability is not broken for any microscopic structure and PES, consistent with our previous study. 
Meanwhile, Eq.~\eqref{eq:J} is not generally satisfied when $\phi$ is a nonlinear map. 
With these considerations, bidirectional stability should be broken due to nonlinear character of thermodynamic average, where we would like to formulate $B_k$ without using any thermodynamic information such as $u_i$, but using $D$ derived only from geometric character of lattice. 

With these preparations, before formulating $B_k$ for systems with $f$-degree of freedom, we first show derivation with $f=2$ in order to essentially capture how $B_k$ is dominated by configurational geometry.  
Since at thermodynamic limit of $N\to \infty$ ($N$ is the number of lattice points in the system), $\Lambda$ becomes diagonal (i.e., $r_{12}\simeq 0$, which immediately leads to 
\begin{widetext}
\begin{eqnarray}
\label{eq:JJ0}
\frac{J}{J^{\left(0\right)}} - 1 &&\simeq -\frac{1}{r_{11}}\left(u_1 d_{111}+u_2 d_{112}\right) -\frac{1}{r_{22}}\left(u_1 d_{122}+u_2 d_{222}\right) + \frac{1}{r_{11}r_{22}}\left(u_1 d_{111}+u_2 d_{112}\right)\left(u_1 d_{122}+u_2 d_{222}\right) \nonumber \\
&& -\frac{\left(r_{12}\right)^2}{r_{11}r_{22}} + \frac{2r_{12}}{r_{11}r_{22}}\left(u_1d_{112} + u_2d_{122}\right) - \frac{1}{r_{11}r_{22}}\left(u_1d_{112}+u_2d_{122	}\right)^2.
\end{eqnarray} 
\end{widetext}
In order to describe Eq.~\eqref{eq:JJ0} in terms of information about anharmonicity, $D$, we should apply the transformation of $u_l=-Q_l/r_{ll}$:
\begin{widetext}
\begin{eqnarray}
\label{eq:JJ01}
\frac{J}{J^{\left(0\right)}} - 1 &&= \frac{Q_1 d_{111}}{\left(r_{11}\right)^2} + \frac{Q_2d_{222}}{\left(r_{22}\right)^2} + \frac{Q_1d_{122}+Q_2d_{112}}{r_{11}r_{22}} \nonumber \\
&&+ \frac{\left( Q_1\right)^2 \left\{ d_{111}d_{122} -  \left(d_{112}\right)^2\right\} }{\left(r_{11}\right)^3 r_{22}} + \frac{Q_1Q_2\left(d_{111}d_{222}+d_{112}d_{122} - 2 d_{112}d_{122}\right)}{\left(r_{11}r_{22}\right)^2} +\frac{\left( Q_2\right)^2 \left\{ d_{112}d_{222} -  \left(d_{122}\right)^2\right\} }{r_{11}\left(r_{22}\right)^3}.
\end{eqnarray}
\end{widetext}
Corresponding to Eq.~\eqref{eq:taylor}, we have derived that anharmonicity along chosen coordination i, $D_i$, is expressed as
\begin{eqnarray}
D_i \simeq \frac{1}{2}\sum_{j=1}^f \sum_{k=1}^f \frac{Q_j Q_k d_{ijk}}{r_{jj}r_{kk}}.
\end{eqnarray}
When we consider that anharmonicity $D$ not as scalar, but as vector field $\mathbf{D}=\left(D_1, D_2, \cdots, D_f\right)$, Eq.~\eqref{eq:JJ01} can be rewritten as 
\begin{eqnarray}
\frac{J}{J^{\left(0\right)}} = 1+ \mathrm{div} \mathbf{D} + J\left( \frac{\partial \mathbf{D}}{\partial \mathbf{Q}} \right),
\end{eqnarray}
where $\mathbf{Q}=\left(Q_1,\cdots, Q_f\right)$. 
Therefore, bidirectional stability breaking at $Q^{\left(m\right)}$ is given by 
\begin{eqnarray}
\label{eq:B2}
B^{\left(m\right)} \simeq \log \left| 1+\mathrm{div}\mathbf{D}^{\left(m\right)} + J\left( \frac{\partial \mathbf{D}^{\left(m\right)}}{\partial \mathbf{Q}^{\left(m\right)}} \right)  \right|
\end{eqnarray}
for $f=2$. 
Note that for the special case of $f=1$, Eq.~\eqref{eq:B2} is simplified to
\begin{eqnarray}
B^{\left(m\right)} \simeq \log \left| 1+\mathrm{div}\mathbf{D}^{\left(m\right)}   \right|. 
\end{eqnarray}
Eq.~\eqref{eq:B2} clearly indicates that bidirectional stability breaking is due to the sum of linear change of nonlinear character in $\phi$, $\mathrm{div}\mathbf{D}$,  and of nonlinear change in $\phi$, 
$J\left(\partial \mathbf{D}/\partial \mathbf{Q}\right)$. The most important point here is that the breaking can be expressed only by information about underlying geometry of lattice, without using any thermodynamic information such as energy or temperature. 
For systems with $f > 2$, we can show that by recursively applying cofactor expantion to the Jacobian, Eq.~\eqref{eq:B2} is generalized to
\begin{widetext}
\begin{eqnarray}
B^{\left(m\right)} \simeq \log \left| 1+\mathrm{div}\mathbf{D}^{\left(m\right)} + \sum_{\mathbf{F}} J_{\mathbf{F}}\left( \frac{\partial \mathbf{D}^{\left(m\right)}}{\partial \mathbf{Q}^{\left(m\right)}} \right)+ J\left( \frac{\partial \mathbf{D}^{\left(m\right)}}{\partial \mathbf{Q}^{\left(m\right)}} \right)  \right|,
\end{eqnarray}
\end{widetext}
where $\sum_{\mathbf{F}}$ denotes taking summation over Jacobian for all possible subspaces for $J\left(\partial \mathbf{D}^{\left(m\right)}/\partial \mathbf{Q}^{\left(m\right)}\right)$. 
Therefore, for general system with $f$ degree of freedoms, interpretation of bidirectional stability breaking $B^{\left(m\right)}$ in terms of vector field of anharmonicity, $\mathbf{D}$ is the same as in the case of $f=2$.

We finally note that since the breaking can be characterized by divergence and Jacobian, the following trajectory of introduced discrete dynamical system in future study becomes important to geometrically capture the breaking character:
\begin{eqnarray}
Q^{\left(t+1\right)} = \left( \phi_{\textrm{th}}\left( \beta \right)\circ\Lambda^{-1}\left( \beta \right)  \right) \cdot Q^{\left(t\right)},
\end{eqnarray}
where $t$ takes non-negative integer value starting from zero. From the definition of anharmonicity and the used transformation of $u_l=-Q_l/r_{ll}$, we have to analyze the vector field $\mathbf{D}$ in terms of endpoint of the discrete trajectry, since we see the stability breaking for $Q^{\left(t+1\right)}$ at $Q^{\left(t\right)}$.


\section{Conclusions}
We quantitatively formulate bidirectinal stability breaking of thermodynamic average in terms of vector field of anharmonicity in the structural degree of freedom, $D$. 
The derived formulation does not require any thermodynamic information, indicating that purely geometric character of underlying lattice plays essential role to determine the stability breaking. 
We find that the breaking is mainly due to divergence and Jacobian of $D$ in configuration space, indicating that linear and nonlinear change is nonlinear map character of thermodynamic average 
enhance the bidirectinal stability breaking.

\section{Acknowledgement}
This work was supported by a Grant-in-Aid for Scientific Research (16K06704) from the MEXT of Japan, Research Grant from Hitachi Metals$\cdot$Materials Science Foundation, and Advanced Low Carbon Technology Research and Development Program of the Japan Science and Technology Agency (JST).

\end{document}